\documentclass[aps,twocolumn]{revtex4}
\usepackage{bm}
\usepackage{graphicx}

\begin{document}

\title{Phase diagram and complexity of mode-locked lasers: from order to disorder
}

\author{L. Leuzzi$^{1,3}$, C. Conti$^{2,3}$,
V. Folli$^{3}$, L. Angelani$^{1,3}$, G. Ruocco$^{2,3}$} 

\affiliation{ $^1$Research center SMC INFM-CNR, c/o Universit\`a di
Roma ``Sapienza,'' I-00185, Roma, Italy\\ $^2$Research center Soft
INFM-CNR, c/o Universit\`a di Roma ``Sapienza,'' I-00185, Roma, Italy
\\ $^3$Dipartimento di Fisica, Universit\`a di Roma ``Sapienza,''
I-00185, Roma, Italy}

\begin{abstract}
We investigate mode-locking processes in lasers displaying a variable
degree of structural randomness, from standard optical cavities to
multiple-scattering media.  By employing methods mutuated from
spin-glass theory, we analyze the mean-field Hamiltonian and derive a
phase-diagram in terms of the pumping rate and the degree of disorder.
Three phases are found: i) paramagnetic, corresponding to a noisy
continuous wave emission, ii) ferromagnetic, that describes the
standard passive mode-locking, and iii) the spin-glass in which the
phases of the electromagnetic field are frozen in a exponentially
large number of configurations. The way the mode-locking threshold is
affected by the amount of disorder is quantified.  The results are also
relevant for other physical systems displaying a random Hamiltonian,
like Bose-Einstein condensates and nonlinear optical beams.
\end{abstract}

\maketitle

\begin{figure}[t!]
\includegraphics[width=.49\textwidth]{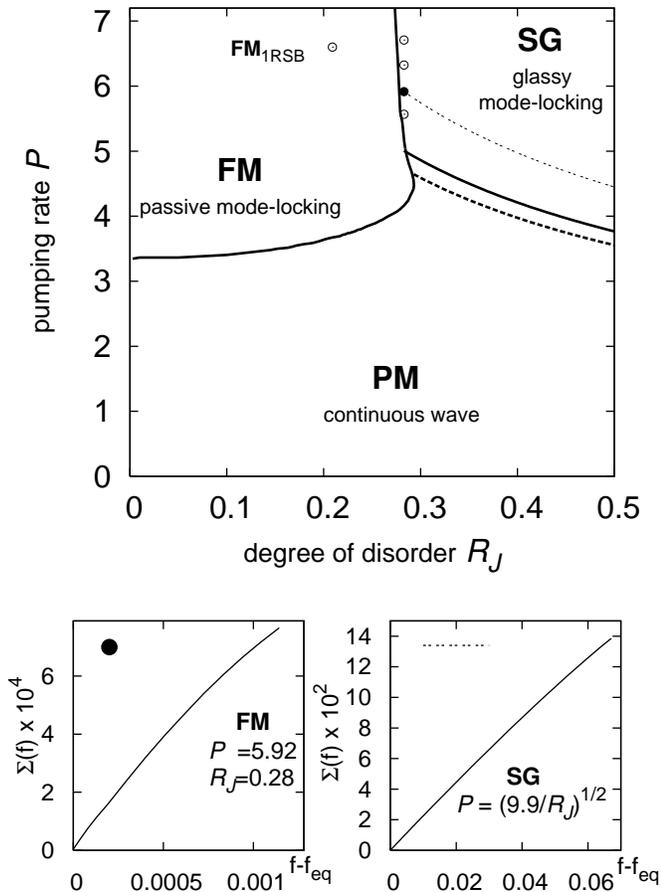}
\protect\caption{Phase diagram in the plane $(R_J,{\cal P})$.  Three
phases are found: PM (low ${\cal P}$), FM (high ${\cal P}$/weak
disorder) and SG (high ${\cal P}$/strong disorder).  The full line
is thermodynamic transitions, the dashed line represents the dynamic
PM/SG transition.
 The transition line to the FM phase (both from PM and from SG) are
obtained using the RS approximation.  The circles are exact 1RSB FM
solutions. In the insets complexity vs. free energy curves are plotted
in the SG phase (right inset, at $\bar \beta=9.9$, $R_J$ independent;
in the main plot: $p=\sqrt{9.9/R_J}$, tiny-dashed line) and in the FM
phase next to the SG/FM transition (left inset, at $R_J=0.28$,
${\cal P}=5.92$; full circle in the main plot).  The latter is two order of
magnitude smaller.}
\label{fig:PhDi}
\end{figure}

\begin{figure}[t!]
 \includegraphics[width=.49\textwidth]{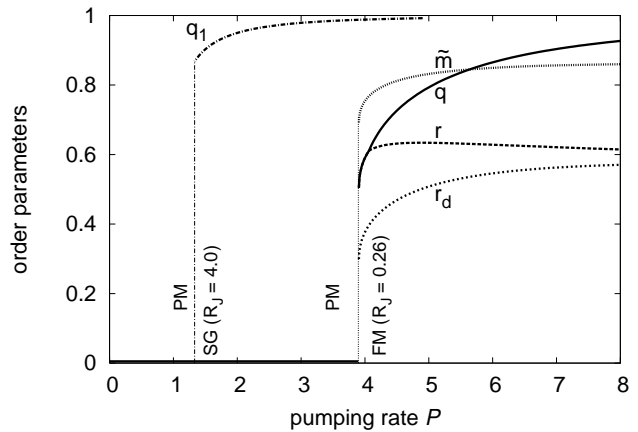}
 \protect\caption{Discontinuity of the order parameters at the
 transition point in the pumping rate $p$.
Left: jump in $q_1$ at the PM/SG
($R_J=4.0$). Right: Discontinuous  $\tilde m$,
 $r$, $q$, $r_d$ at the PM/FM transition ($R_J=0.26$).
In the FM phase the  thermodynamics is computed in the
 RS approximation ($q_1=q_0=q$, $r_1=r_0=r$).}
 \label{fig:op}
 \end{figure}

\vskip 5 cm The number of different disciplines converging in the
field of disordered lasers is impressive; random lasers embrace
photonics \cite{Cao03r}, wave-transport and localization
\cite{John96,Lubatsch05,Wiersma95,lepri:07}, spin-glass theory
\cite{Angelani06}, random-matrices \cite{Beenakker:98}, soft- and
bio-matter \cite{Zhang:95,Polson04,Siddique:96}, nonlinear and quantum
physics
\cite{Florescu04b,Deych05,Tureci:08,Patra02,Hackenbroich:01,Berger:97}.
Notwithstanding the several theoretical and experimental
investigations, as recently reviewed in \cite{Wiersma:08,Cao03r}, many
are the open issues and, certainly, the development of the field
cannot be compared with that of standard lasers (SL) (i.e., those not
displaying structural disorder), even if the first theoretical
prediction of RL \cite{Ambartsumyan:66} is dated not very far from the
first theoretical work on SL \cite{Schawlow:58}.  In this respect, a
comprehensive theory of stimulated emission able to range from ordered
to disordered lasers will be certainly an important step. This theory
should be able to account for the strength of disorder as a parameter
and should predict specific regimes attainable in SL and RL.  At the
moment such a theory is not avalaible, and the literature dealing with
the two kinds of lasers is still largely disjoint. Furthermore,
nano-structured lasers unavoidably display some degree of disorder due
to fabrication tolerances; hence understanding the effect of
randomness in the light emission has important practical relevance.
The main question addressed in this work is the following: consider a
SL operating in mode-locking (ML) and progressively increase the
amount of structural disorder, at which value one should expect that
the mode-locking is frustrated? Which kind of states are expected?
\\
\indent In this manuscript we derive a self-consistent theoretical
approach to ML in ordered and disordered lasers; we make use of all
the machinery inherited from spin-glass theory \cite{ParisiBook}, at
the moment the only mathematical technique enabling to fully account
for different degree of randomness and for several other phenomena
expected in complex systems.  We report on a phase-diagram for ML
in lasers 
unveiling the interplay between randomness and nonlinearity. We
discover the existence of different phases characterized by a
non-trivial configurational entropy, or {\it complexity}, which
measures the number of energetically equivalent ML states.  
\\ \indent Previous theoretical work, which was based on thermodynamic
approaches, has dealt with the two regimes: the ordered case where the
ML is demonstrated to be given by a ferromagnetic-like transition
\cite{HakenBook,Weill05,angelani:07} and the completely disordered
limit \cite{Angelani06}.  The former case being relevant to fiber or
dye/solid-state standard passive ML \cite{Haus:00}; the latter being
more oriented on stimulated emission in the presence of multiple
scattering \cite{Wiersma:08,Cao03}.  Our analysis unifies the
scenario, by showing that it is possible to treat in a single
systematic way ML processes in the presence of an arbitrary degree of
disorder. The study is based on the following steps: 1) perform a
statistical average of the disordered free-energy of the system; 2)
identify the order parameters; 3) determine the thermodynamic
free-energy; 4) build the phase-diagram; 5) compute the complexity.
\\ \indent Our main finding can be summarized as follows: the
self-starting ML process maintains its standard ``ferromagnetic''
character (i.e., an abrupt transition from a continuos wave operation
to a pulsed regime) as far as the structural fluctuactions are
sensitively smaller than the average value of the mode coupling
coefficients, although the pumping threshold for passive ML grows with
the amount of disorder. Conversely, for large disorder the transition
acquires a glassy character, the pumping threshold becomes independent
of the ``disorder to order'' ratio and the complexity is not
vanishing; this implies that there exists a large number of ML
states distributed in a given free energy interval and large
fluctuactions from pulse to pulse are expected.  Also an intermediate
regime occurs, where the system mantains
a ferromagnetic behavior but a non-zero complexity is found.
\\ 
\indent {\it
Leading model ---} Under general hypotheses (discussed with details in
\cite{Angelani06long,Gordon03}) the mode-phase dynamics of a RL can be
cast into a dynamical problem with a random Hamiltonian
\begin{equation}
{\cal H}_J[\phi]=
-\sum_{\footnotesize{\begin{array}{c} i_1<\!i_2, j_1<\!j_2
\\ 
i_1<\!j_1 \end{array}}}^{N}J_{\bf{i}\bf{j}}
\cos(\phi_{i_1}+\phi_{i_2}-\phi_{j_1}-\phi_{j_2})
\label{f:Ham}
\end{equation}
where $N$ is the number of electromagnetic modes, $\phi_i$ are their
phases, ${\bf i}=(i_1,i_2)$ and the quenched couplings have a Gaussian
distribution with ${\overline{J_{\bf i j}}}=J_0/N^3$ and ${\overline
{(J_{\bf ij}-{\overline{J_{\bf ij}}})^2}}=\sigma^2_J/N^3$. The overbar
denotes the average over the disorder, which is quantified by the
parameter $R_J\equiv \sigma_J/J_0$ that is the ratio between the
standard deviation of the distribution of the coupling coefficients
$J_{\bf ij}$ and their mean.  The limit $R_J\rightarrow 0$
($R_J\rightarrow\infty$) corresponds to the ordered (disordered) case.
The other relevant parameter is the normalized pumping threshold for
mode-locking \cite{Angelani06long} ${\cal P}$, that is connected to the
thermodynamic inverse temperature $\beta$ and to the interactions by
the relationship ${\cal P}=\sqrt{\beta J_0}=\sqrt{\bar \beta/R_J}$,
where $\bar\beta=\beta\sigma_J$.
In our units, when $R_J\rightarrow0$,
 $p=p_{ord}\cong 3.339$ (see Fig.\ref{fig:PhDi}), in agreement with the
 ordered case \cite{Angelani06}\footnote{A factor of $8$ has to be
 considered because of the overcounting of terms in the Hamiltonian of
 the model studied in Ref. \cite{Angelani06} with respect to
 Eq. (\ref{f:Ham}). This factor can be absorbed into the temperature
 yielding the pumping threshold $p_{\rm ord}=\sqrt{8/T_0}\cong 3.34$,
 where $T_0\cong 0.717$ is the temperatute at which the FM phase first
 appears in complete absence of disorder.}; as detailed below, the
 deviation from this value quantifies an increase of the standard
 mode-locking threshold $p_{ord}$ due to disorder.  The specific value
 for $p_{ord}$ will depend on the class of lasers under consideration
 (e.g., a fiber loop laser or a random laser with paint pigments), but
 the trend of the passive mode-locking threshold with the strendth of
 disorder $R_J$ in Fig.\ref{fig:PhDi} (FM/PM transition line, see
 below) has a universal character.
\\ \indent {\it Averaged free-energy ---} The average free energy is
calculated by the replica trick \cite{ParisiBook}.  By considering $n$
copies of the system, Eq. (\ref{f:Ham}), the free energy averaged over
the disorder can be computed as
\begin{equation}
\beta \Phi=-\frac{1}{N}~{\overline {\log Z_J}}=-\frac{1}{N} \lim_{n\to
0}\frac{{\overline {Z^n_J}}-1}{n} \ .
\end{equation}  
The thus {\em replicated} partition
function, ${\overline Z^n_J}$, takes the form
\begin{eqnarray}
{\overline {Z^n}_J}\propto \int {\cal D}{\bm X}
\exp\left[-n N G({\bm X})\right]
\sim e^{-n N G({\bm X}_{\rm SP})}
\label{f:Zrep}
\end{eqnarray}
where ${\bm X}$ denotes the set of all the order parameters and the integral is
evaluated by means of the saddle point approximation (valid for large
$N$).
\\ \indent Spin-glass systems described by more-than-two-body
interactions, cf. Eq. (\ref{f:Ham}), are known to have low temperature
phases whose correct thermodynamic description is provided by the
so-called one step Replica Symmetry Breaking (1RSB) Ansatz
\cite{Gardner85,Crisanti92}.  Under this Ansatz, taking the $n\to 0$
limit, the free energy functional reads \footnote{In
Ref. \cite{Angelani06} the standard deviation of the quenched disorder
was fixed at $\sigma^2_J=8$. In this work the adimensional temperature
introduced there is computed in units of $\sigma_J$.  }
\begin{eqnarray}
\label{f:repPhi_1rsb}
&&\beta \Phi 
=-\frac{\bar\beta R_J}{8}|{\tilde m}|^4
-\frac{{\bar\beta}^2}{32}\Bigl[
1
-(1-m)\left(|q_1|^4+|r_1|^4\right)
\\
\nonumber
&&\quad -m\left(|q_0|^4+|r_0|^4\right)+|{r_d}|^2
\Bigr]
 -\Re\Big[\frac{1-m}{2}\left(\bar\lambda_1q_1+\bar\mu_1r_1\right)
\\
\nonumber
&&\quad+\frac{m}{2}\left(\bar\lambda_0q_0+\bar\mu_0r_0\right)
-{\bar \mu_d} {r_d} - {\bar \nu}{\tilde m}\Bigr]+ \frac{\lambda_1^R}{2}
 \\
\nonumber
&&\quad -\frac{1}{m}\int {\cal D}[\bm{0}]\log\int{\cal D}[\bm{1}]
\left[\int_0^{2\pi}\!d\phi~\exp{\cal L}(\phi; \bm{0},\bm{1})
\right]^m
\end{eqnarray}
where $\bm{0}=\{x_0,\zeta_0^R,\zeta_0^I\}$, 
$\bm{1}=\{x_1,\zeta_1^R,\zeta_1^I\}$, 
${\cal D}[\bm{a}]$ is the product of three Normal distributions
and
\begin{eqnarray}
\nonumber
&&{\cal L}(\phi;\bm{0},\bm{1})\equiv\Re\Bigl\{e^{i\phi}\Bigl[
\bar\zeta_1\sqrt{\Delta \lambda^R-|\Delta \mu|}+
\bar\zeta_0\sqrt{\lambda_0^R-|\mu_0|}+
\\
\label{f:calL}
&&\quad x_1\sqrt{2\Delta \bar\mu}
+x_0\sqrt{2\bar\mu_0}
+\bar\nu\Bigr]
+e^{2i\phi}\left(\bar \mu_d-\frac{\bar\mu_1}{2}\right)
\Bigr\}
\end{eqnarray}
with $\Delta\lambda=\lambda_1-\lambda_0$, $\Delta \mu=\mu_1-\mu_0$.
For later convenience we define the following averages over the action
$e^{\cal L}$, cf. Eq. (\ref{f:calL}): $c_{\cal L}\equiv \langle
\cos\phi\rangle_{\cal L}$, $s_{\cal L}\equiv \langle
\sin\phi\rangle_{\cal L}$.
\\
\indent
The values of the order parameters $\lambda_{0,1}, \mu_{0,1}, \mu_d$ and $\nu$
are yielded by
\begin{eqnarray}
&&\lambda_{0,1}=\frac{{\bar\beta}^2}{4} \left(q_{0,1}\right)^3~
;\quad 
\mu_{0,1}=\frac{{\bar\beta}^2}{4}|r_{0,1}|^2~ r_{0,1}
\\
&&\mu_d= \frac{{\bar\beta}^2}{8}|r_d|^2{r_d}~~
;\qquad \nu=\frac{\bar\beta R_J}{2}|\tilde m|^2{\tilde m}
\end{eqnarray}
The remaining parameters 
 are obtained by solving the
self-consistency equations:
\begin{eqnarray}
\label{f:speq_q1_R}
&&\hspace*{-4mm}
q_1=\langle \langle c_{\cal L}^2 \rangle_{m}\rangle_{\bf 0}+
\langle \langle  s_{\cal L}^2\rangle_{m}\rangle_{\bf 0}
\\
\label{f:speq_q0_R}
&&\hspace*{-4mm}
q_0=\langle \langle c_{\cal L} \rangle_{m}^2\rangle_{\bf 0}+
\langle \langle  s_{\cal L}\rangle_{m}^2\rangle_{\bf 0}
\\
\label{f:speq_r1R}
&&\hspace*{-4mm}
r_1= \langle \langle c_{\cal L}^2 \rangle_{m}\rangle_{\bf 0}
-\langle \langle s_{\cal L}^2 \rangle_{m}\rangle_{\bf 0}
+2 \rm i \langle \langle c_{\cal L} s_{\cal L} \rangle_{m}\rangle_{\bf 0}
\\
\label{f:speq_r0R}
&&\hspace*{-4mm}
r_0=\langle \langle c_{\cal L} \rangle_{m}^2\rangle_{\bf 0}
-\langle \langle s_{\cal L} \rangle_{m}^2\rangle_{\bf 0}
+2 \rm i \langle \langle c_{\cal L} \rangle_{m}\rangle_{\bf 0}
\langle \langle  s_{\cal L} \rangle_{m}\rangle_{\bf 0}~~
\\
\label{f:speq_rd}
&&\hspace*{-4mm}
r_d=\langle \langle \langle
e^{2\rm i \phi}
\rangle_{\cal L}\rangle_{m}\rangle_{\bf 0};
\qquad\tilde m=\langle \langle 
\langle e^{\rm i \phi}
\rangle_{\cal L}\rangle_{m}\rangle_{\bf 0}
\end{eqnarray}
where
\begin{eqnarray}
\langle(\ldots )\rangle_m&\equiv&
\frac{\int{\cal D}[\bm{1}](\ldots )\left[\int_0^{2\pi}\!
d\phi~e^{{\cal L}(\phi;\bm{0},\bm{1})}
\right]^m}
{\int{\cal D}[\bm{1}]\left[\int_0^{2\pi}\!d\phi~e^{{\cal L}
(\phi;\bm{0},\bm{1})}\right]^m}
\\
\langle(\ldots )\rangle_{\bf 0}
&\equiv&
\int{\cal D}[\bm{0}](\ldots )
\end{eqnarray}
The overlap parameters $q_{0,1}$  are real-valued,
whereas $r_{0,1}, r_d$ and ${\tilde m}$ 
are complex. ``One step'' parameters $X_{0,1}$ ($X=q,r$)  enter with a
probability distribution that can be parametrized by the so-called
replica symmetry breaking parameter $m$, such that, e.g.,
$P(X)=m~\delta(X-X_0)+(1-m)\delta(X-X_1)$.
\\ \indent {\it Complexity ---} The resulting eleven independent
parameters that can be evaluated by solving
Eqs. (\ref{f:speq_q1_R})-(\ref{f:speq_rd}) must be combined with a further
equation for the parameter $m$.  This is stricly linked to the
expression for the {\em complexity} function of the system, i.e., the
average logarithm of the number of states of the system present at a
given free energy level $f$. The complexity can be computed, e.g., 
as the Legendre transform of the replicated free
energy, Eq. (\ref{f:repPhi_1rsb}):
\begin{eqnarray}
\Sigma 
\label{f:Sigma}
&=& \min_m \left[-\beta m \Phi(m)+\beta m f\right]
=  \beta m^2\frac{\partial\Phi}{\partial m}
\\
&=&\frac{3}{4}\beta^2m^2
\left(|q_1|^4+|r_1|^4-|q_0|^4-|r_0|^4\right)
\nonumber
\\
\nonumber
&& +\int {\cal D}[\bm{0}]\log\int{\cal D}[\bm{1}]
\left[\int_0^{2\pi}\!\!\!\!d\phi~\exp{\cal L}(\phi; \bm{0},\bm{1})
\right]^m
\\
&& -m\int {\cal D}[\bm{0}]\langle\log
\int_0^{2\pi}\!\!\!\!d\phi~\exp{\cal L}(\phi; \bm{0},\bm{1})
\rangle_m
\nonumber
\end{eqnarray}
where the single state free energy $f=\partial (m\Phi)/\partial m$ is
conjugated to $m$. Since the above expression is proportional to
$\partial \Phi/\partial m$, equating $\Sigma=0$ 
provides the missing equation to determine the order parameters
values.  
\\ \indent {\it Phase-diagram ---} By varying the normalized pumping
rate ${\cal P}$ and degree of disorder $R_J$, we find three different
phases, as shown in Fig.  \ref{fig:PhDi}.
For low ${\cal P}$ and $R_J$ the only phase present is completely
disordered: all order parameters are zero and we have a ``paramagnet''
(PM); the laser emission is expected to be given by a noisy continuous
wave emission, and all the mode-phases are uncorrelated.  The PM exists
everywhere in the whole plane ($R_J$, ${\cal P}$), becoming
thermodynamically subdominant as ${\cal P}$ and $R_J$ increase and other
phases arise.  For large disorder, as the pumping rate is increased, a
discontinuous transition occurs from the PM to a spin-glass (SG) phase
in which the phases $\phi$ are frozen at given values, though not
displaying any ordered pattern in space.  First, along the line
${\cal P}_d=\sqrt{\bar\beta_d/R_J}$ ($\bar\beta_d=6.322$), a dynamic
transition occurs (dashed line). Indeed, the lifetime of metastable
states is infinite in the mean-field model and the dynamics gets stuck
in the highest lying excited states.  The thermodynamic (lowest
energy) state is, however, still PM. Accross the full line
(cf. Fig. \ref{fig:PhDi}) ${\cal P}_s=\sqrt{\bar \beta_s/R_J}$
($\bar\beta_s =7.094$) a true thermodynamic phase transition to the SG
phase occurs.
The order parameter $q_1$, else called
Edwards-Anderson parameter $q_{\rm EA}$ \cite{Edwards75}
discontinuously jumps at the transition to a non-zero value
$q_{1}>q_0=0$ and $\tilde m=r_0=r_1=r_d=0$ (see Fig. \ref{fig:op}).
The SG phase exists for any value of $R_J$ and $\bar \beta> \bar \beta_s$.
However, for small $R_J$, a ferromagnetic (FM) phase turns
out to dominate over the SG and the PM phases. 
The transition line PM/FM is the
standard passive ML threshold (see e.g. \cite{Gordon02,Gordon03}) 
and from Fig.\ref{fig:PhDi} we see that
it takes place at growing pumping rates for increasing disorder,
as far as the SG is not encountered.
\\ \indent To precisely describe the FM phase in the 1RSB Ansatz we
have to solve eleven coupled integral equations
[Eqs. (\ref{f:speq_q1_R})-(\ref{f:speq_rd}) and $\Sigma=0$,
cf. Eq. (\ref{f:Sigma})].  In the region where this FM$_{\rm 1rsb}$
phase is the thermodynamically dominant solution, however, the PM and
the SG solutions also satisfy the same set of equations.  Starting the
iterative resolution with randomly chosen initial conditions,
determining the SG/FM$_{\rm 1rsb}$ and the PM/FM$_{\rm 1rsb}$
transition lines becomes numerically very demanding.  An approximation
is obtained by considering the Replica Symmetric (RS) solution for the
FM phase (FM$_{\rm rs}$). This reduces the number of independent
parameters to five ($q_1=q_0$, $r_1^{R,I}=r_0^{R,I}$, $r_d$ and
$\tilde m$). The corresponding transition line is shown in
Fig. \ref{fig:PhDi}.  The exact FM phase is provided by a 1RSB
solution and some sampled points are represented by the circles in the
phase diagram.
\\ \indent The 1RSB ansatz also enables to determine the not-vanishing
extensive complexity, which signals the presence of a huge quantity of
excited states with respect to ground states. This also implies the
occurrence of dynamic transitions besides the thermodynamic one, as
anticipated.  These take place between PM and SG [where the state
structure always displays a non-trivial $\Sigma(f)$ for any $\bar
\beta>\bar \beta_d$] and in the FM phase, though the magnitude of
$\Sigma$ turns out to be smaller.
In the left inset of
Fig. \ref{fig:PhDi} we show, e.g., $\Sigma(f)$ in the FM$_{\rm 1RSB}$
phase at $(R_J,{\cal P})=(0.28,5.92)$.  This has to be
compared with the SG complexity at the same temperature (right inset of
Fig. \ref{fig:PhDi}) that is sensitively larger and does not depend on
the ``disorder/order ratio'': the maximum complexity drops of about
two order of magnitude at the SG/FM$_{\rm 1rsb}$ transition, thus
unveiling a corresponding {\em high to low complexity transition}.
\\ \indent {\it Conclusion ---} We have reported on what we believe to
be the first comprehensive theoretical treatment of ML processes in
lasers with structural disorder.  Our approach enables to take into
account an arbitrary strength of randomness in the system and results
into a phase-diagram for random lasers. In this diagram, a
ferromagnetic-like (standard passive mode-locking) and glassy-like
phases (``glassy'' mode-locking) are present and determined by the
pumping rate and by the degree of structural disorder. 
The way the self-starting passive mode-locking
threshold is affected by the latter quantity has been quantified.
This result can be experimentally tested in a variety of different
physical systems, from laser powders to standard laser cavities, 
and is relevant for any nonlinear interaction process of
several oscillators in the presence of disorder, also including
Bose-Einstein condensation.  The phase-diagram reveals a rich variety
of different phases, and even if only those thermodynamically favored
are reported above, the dynamics of the mode-locking process (and the
corresponding synchronization of the nonlinearly coupled resonances)
is expected to be strongly affected by the existence of several
valleys in the free energy (described by a not vanishing {\it
complexity}). In this respect, lasing in disordered system is an
important framework for investigating out of equilibrium dynamical
systems, including quantum effects.
 
{\it Acknowledgments ---}
The research leading to these results has received funding from
the European Research Council under the European Community's Seventh Framework
Program (FP7/2007-2013)/ERC grant agreement n. 201766.

\end{document}